\documentclass[pre,11pt]{revtex4}
\usepackage{bm}


\def\al{\alpha}
\def\del{\delta}
\def\Del{\Delta}

\def\be{\begin{equation}}
\def\ee{\end{equation}}
\def\bea{\begin{eqnarray}}

\def\eea{\end{eqnarray}}
\def\la{\label}
\def\bsea{\begin{subeqnarray}}
\def\esea{\end{subeqnarray}}
\def\u{\underline}

\begin{document}
\title{Structural Fluctuation of Protein in Water around Its Native State: 
A New Statistical Mechanics Formulation}
\author{Bongsoo Kim${}^ 1$ and Fumio Hirata${}^ 2$}
\affiliation{${}^ 1$ Department of Physics and Institute for Soft and Bio Matter Science, Changwon National University, 
Changwon 641-773, Korea\\
${}^ 2$ College of Life Sciences, Ritsumeikan University, Kusatsu, Shiga 525-8577, Japan
}
\date{\today}
\begin{abstract}
A new statistical mechanics formulation of characterizing the structural fluctuation of protein correlated with 
that of water is presented based on the generalized Langevin equation and the 3D-RISM/RISM theory of molecular liquids.
The displacement vector of atom positions and their conjugated momentum, are chosen for 
the dynamic variables for protein, while the density fields of atoms and their momentum fields are chosen
for water. Projection of other degrees of freedom onto those dynamic variables using the standard projection
operator method produces essentially two equations which describe the time evolution of fluctuation concerning
the density field of solvent and the conformation of protein around an equilibrium state, which are coupled with
each other. The equation concerning the protein dynamics is formally akin to that of the coupled Langevin 
oscillators, and is a generalization of the latter, to atomic level.
The most intriguing feature of the new equation is that it contains the variance-covariance matrix as the 
"Hessian" term describing the "force" restoring an equilibrium conformation, which is the second moment
of the fluctuation of atom positions. The "Hessian" matrix is naturally identified as the second derivative of 
the free energy surface around the equilibrium.
A method to evaluate the Hessian matrix based on the 3D-RISM/RISM theory is proposed. Proposed also is 
an application of the present formulation to the molecular recognition, in which the conformational fluctuation of protein 
around its native state becomes an important factor as exemplified by so called "induced fitting".
\end{abstract}

\maketitle
\section{Introduction}
\setcounter{equation}{0}
Structural fluctuation of protein around its native state plays essential roles in a variety of processes in which the biomolecule performs its intrinsic function 
\cite{protein}. For example, so called "gating" mechanisms of ion channels are regulated by the structural fluctuation of amino-acid residues consisting the gate region of the channel. Molecular recognition such as the formation of an enzyme-substrate complex in an enzymatic reaction is controlled often by structural fluctuation of protein. 
Few typical examples of structural fluctuations around a native conformation of protein, related to function, are "breathing"\cite{breathing},  "hinge-bending"\cite{hinge},
 and "arm-rotating" motions \cite{arm}. 
Those motions are {\em collective} in nature involving many atoms moving in the same direction. Those structural fluctuation associated with protein functions, whether it's large or small, stays within its native conformation, and does not induce global conformational change such as denaturing, with few exceptions exemplified by {\em intrinsically disordered protein} \cite{idp}.  
   
 In actual biological processes, solvent plays vital roles both in the equilibrium and in fluctuation of protein \cite{mfeig}. It may not be necessary to spend many words for emphasizing the crucial role played by solvent for stabilizing or destabilizing native structure of protein, such as the hydrophobic interaction and hydrogen bonds. Here, let us consider roles played by water in fluctuation of protein around its native conformation, associated with recognition of a ligand by protein. The process is primarily a thermodynamic process, governed by the free energy difference between the two states before and after the recognition. It is obvious that water plays crucial role in the thermodynamics, since the equilibrium structures are determined by the free energies including the excess chemical potential or the solvation free energy of water. However, it is not the only role of water in the process. 
Water actually regulates the kinetic pathway of the process as well by controlling the structural fluctuation of amino-acid residues consisting the active site. An example of such processes is a mouth-like motion of amino-acid residues. The open-and-close motion of the mouth is driven not only by the direct force acting among atoms in protein, but by that originated from the solvent induced force which is in turn caused by the fluctuation in the solvation free energy, or the non-equilibrium free energy. In an actual biomolecular process, such conformational change around the native state is induced often by some perturbation upon amino-acid residues around the active site, for example, binding of a ligand. However, response to the perturbation should be linear, because the protein recovers its native conformation upon removing the perturbation \cite{ikeguchi}. 
   
 It is not surprising that considerable efforts have been devoted to clarify the conformational fluctuation of protein theoretically, which has started at the end of the last century based on the molecular mechanics or dynamics. One of earliest attempts was to relate the structural fluctuation to the normal mode of protein \cite{noguti_go}.  
 Those works have demonstrated the importance of the collective mode in the fluctuation.
However, those efforts have not provided a realistic physical insight into the dynamics of actual biological processes, since they are concerned with a protein in "vacuum", which obviously cannot describe the fluctuation correlated with that of solvent. 
The principal component analysis involving diagonalization of the variance-covariance matrix of conformational fluctuation, extracted from the molecular dynamics trajectory of a protein in water, has revealed some important aspects of the conjugated fluctuation between a biomolecule and water \cite{kitao}. 
 The lowest frequency mode of fluctuation around a native conformation exhibits an activated transition from a minimum to another minimum in the conformational space, akin to the jump diffusion model of liquids. 
However, the procedure cannot be extended readily to that associated with such a process as ligand binding, because the process is concerned with sampling of large configuration space involving both protein and solvent. It becomes formidable especially when the solvent consists of several chemical components such as the electrolyte solution. 
   
In the present work, we propose a new first-principle approach to treat structural fluctuation of protein conjugated with that of solvent, based on the two theoretical frameworks in the statistical mechanics of liquids, or, the 3D-RISM/RISM 
(Reference Interaction Site Model) theory and the generalized Langevin equation \cite{mori}. 
The 3D-RISM/RISM theory \cite{hirata_book} has proven itself to be capable of predicting the molecular recognition of ligand 
by protein which has a rigid structure \cite{imai}. 
The generalized Langevin equation should be able to describe the fluctuation of a system consisting of protein and solvent around its equilibrium state. Therefore, it is reasonable to expect that the two theories combined together will produce a method which can describe the molecular recognition process by protein, whose structure is fluctuating.   
 An etude of such a theory has been already published by us \cite{kim_hirata} 
  where a much more simplified model, a chain of identical particles in solvent consisting of spherical molecules, was considered. 
The key idea there lies in the choice of dynamic variables. We have chosen four quantities to form a vector in the phase space: the displacement of atom positions in protein from their equilibrium coordinates, the conjugated momentum of those atoms, the fluctuation of the density field of solvent molecules, and their conjugated momentum field or flux. 
A standard treatment of the dynamic variables due to the projection operator method \cite{mori,zwanzig}  gave rise to four equations with respect to the time evolution of those quantities, two for solute and two for solvent, which interplay with each other. 
Most important observation in the results is that the equation of motion concerning the solute dynamics includes the variance-covariance matrix regarding the conformational fluctuation of solute as a "Hessian" or a "force constant" of the "oscillation" or fluctuation.
    Here, we generalize the theory developed in the preceding paper \cite{kim_hirata} 
substantially in order to be able to treat a realistic protein in a realistic solvent such as water.

\section{Projection operator method: summary}
\setcounter{equation}{0}
Since the projection operator method is well-known \cite{balucani,mazenko,reichman}, we here only summarize the general results of the method.
It gives the time evolution equation of a dynamic variable 
${\bf A}(t)$ which is a function of microscopic variables.
Its microscopic time evolution is governed by 
the Liouville operator $i{\cal L}$ whose expression will be given in the next section: 
\be
\frac{d{\bf A}(t)}{dt} \equiv i{\cal L} {\bf A}(t) \equiv 
\{A, {\cal H}\}_{PB} (t)
\la{eqn:2.1}
\ee
where $\{a,b\}_{PB}$ is the Poisson bracket, and ${\cal H}$ 
is the Hamiltonian of the system. 
The formal solution of (\ref{eqn:2.1}) is given by
\be
{\bf A}(t)=e^{i{\cal L}t} {\bf A}(0) \equiv e^{i{\cal L}t} {\bf A}
\la{eqn:2.2}
\ee

Now the projection operator ${\cal P}$ is defined as 
\be
{\cal P} \big(\cdots\big) 
\equiv \big({\bf A}, \cdots \big) \big({\bf A}, {\bf A}\big)^{-1} {\bf A}
\la{eqn:2.3}
\ee
The inner product $({\bf a},{\bf b})$ denotes an average of the canonical 
distribution $\exp\big(-{\cal H}/k_BT\big)$:
\be
({\bf a},{\bf b})\equiv \big<{\bf a}^* {\bf b}\big>
={\cal Z}^{-1} \int d\Gamma \,  {\bf a}^* {\bf b} \exp\big(-{\cal H}(\Gamma)/k_BT\big)
\la{eqn:2.4}
\ee
where $\Gamma$ denotes all microscopic degrees of freedom in the system.
The operator projects out only the 'component' of ${\bf A}$ from the object
$(\cdots)$. Then obviously ${\cal P}{\bf A}={\bf A}$ holds. 
It also has the idempotent property ${\cal P}^2={\cal P}$. 

After projecting  ${\bf A}$-component out of the microscopic degrees of freedom, 
the exact time evolution equation for ${\bf A}(t)$ is given by
\be
\frac{d{\bf A}(t)}{dt}=i{\bf \Omega} \cdot {\bf A}(t)
-\int_0^t ds\, {\bf K}(t-s) \cdot {\bf A}(s) + {\bf f}(t)
\la{eqn:2.5}
\ee
Here the frequency matrix $i{\bf \Omega}$, the memory matrix ${\bf K}(t)$,
 and the fluctuating force vector $ {\bf f}(t)$ are given by 
\bea
i{\bf \Omega}&=&\big({\bf A}, \dot {\bf A}\big) \cdot \big({\bf A}, {\bf A}\big)^{-1},
\nonumber \\
{\bf K}(t)&=& \big( {\bf f}, {\bf f}(t) \big) \cdot \big({\bf A}, {\bf A}\big)^{-1},
\nonumber \\
{\bf f}(t) &=& \exp \Big( t (1-{\cal P}) i{\cal L}\Big) (1-{\cal P}) \dot {\bf A} 
\la{eqn:2.6}
\eea
One can show easily that the fluctuating force ${\bf f}(t)$ 
does not have ${\bf A}$-component, i.e., $({\bf A}, {\bf f}(t))=0$.
Using this feature and the linearity of the equation, we immediately obtain 
the following dynamic equation for 
the auto-correlation function of $ {\bf A}(t)$, $ {\bf C}(t)$
\be
\frac{d{\bf C}(t)}{dt}=i{\bf \Omega} \cdot {\bf C}(t)
-\int_0^t ds\, {\bf K}(t-s) \cdot {\bf C}(s) 
\la{eqn:2.7}
\ee

\section{Generalized Langevin equations for a solute-solvent system} 
\setcounter{equation}{0}
Our main concern here is a protein-water system at infinite dilutions.
However, the formulation is completely general for any solute-solvent system
at infinite dilution. So, in the formulation, we consider 
 a general solute-solvent system. 
 In particular, we consider a solute  molecule consisting of $N_u$ atoms
 immersed in solvent consisting of $N$ molecules, each having $n$ atoms.
The Hamiltonian of the solute-solvent system is then given by
\bea 
{\cal H} &\equiv& {\cal H}_0 + {\cal H}_1+ {\cal H}_{2}, \nonumber \\
{\cal H}_0 &=& \sum_{i=1}^N \sum_{a=1}^n \Big[\frac{{\bf p}^a_i\cdot 
{\bf p}^a_i }{2m_a}+\sum_{j \neq i} \sum_{b \neq a} 
U_0(|{\bf r}^a_i-{\bf r}^b_j|)  \Big] \quad (\mbox{solvent}) \nonumber \\
{\cal H}_1 &=& \sum_{\alpha=1}^{N_u} \Big[
\frac{{\bf P}_{\al} \cdot {\bf P}_{\al}}{2M_{\al}}+\sum_{\beta \neq \al}
U_1(|{\bf R}_{\al}-{\bf R}_{\beta}|)  \Big] \quad (\mbox{solute}) \nonumber \\
{\cal H}_2 &=& \sum_{\al=1}^{N_u} \sum_{i=1}^N \sum_{a=1}^n 
U_{int}(|{\bf R}_{\al}-{\bf r}^a_i|)  \quad (\mbox{solute-solvent})
\la{eqn:3.1}
\eea
where $M_{\al}$ denotes the mass of the $\al$th atom in the solute particle, and $m_a$  the mass of $a$th atom in a solvent molecule.
 The Hamiltonian of the solvent is denoted by ${\cal H}_0$ where ${\bf r}^a_i$ and 
${\bf p}^a_i$ are respectively the position and momentum of $a$th atom in the $i$th molecule of the solvent, 
and $U_0(r^{ab}_{ij})$ ($r^{ab}_{ij}\equiv |{\bf r}^a_i-{\bf r}^b_j|$) 
is the pair potential energy between them. 
${\cal H}_1$ is the Hamiltonian of the $N_u$ solute atoms, and 
${\bf R}_{\al}$ and ${\bf P}_{\al}$ are the position and momentum of the $\al$th solute atom 
(we preserve the Greek indices for denoting the solute atoms),
and $U_{int}(|{\bf R}_{\al}-{\bf r}^a_i|)$  is the interaction potential energy between
the $\al$th solute atom  and the $a$th atom of the $i$th molecule in the solvent. 

The associated Liouville operator $i{\cal L}$ is given by
\bea
i{\cal L} &\equiv& i{\cal L}_0 + i{\cal L}_1, \nonumber \\
i{\cal L}_0 &\equiv & \sum_{i=1}^{N} \sum_{a=1}^n
 \Big[ \frac{1}{m_a} {\bf p}^a_i \cdot 
\frac{\partial}{\partial {\bf r}^a_i}- \sum_{j \neq i} \sum_{b \neq a} 
\frac{\partial U_0(r^{ab}_{ij})}{\partial {\bf r}^a_i} \cdot 
\frac{\partial}{\partial {\bf p}^a_i}
-\sum_{\al=1}^{N_u}\frac{\partial U_{int}(|{\bf R}_{\al}-{\bf r}^a_i|)}
{\partial {\bf r}^a_i} \cdot 
\frac{\partial}{\partial {\bf p}^a_i} \Big] \nonumber \\ 
i{\cal L}_1 &\equiv& \sum_{\al=1}^{N_u} 
\Big[
\frac{{\bf P}_{\al}}{M_{\al}}\cdot 
\frac{\partial}{\partial {\bf R}_{\al}} +{\bf F}_{\al}
 \cdot \frac{\partial}{\partial {\bf P}_{\al}} \Big] 
\la{eqn:3.2}
\eea
where ${\bf F}_{\al} \equiv {\bf F}^{(u)}_{\al}+{\bf F}^{(v)}_{\al}$, and 
$ {\bf F}^{(u)}_{\al}$ is the 
force exerted on the $\al$th solute atom by the other solute atoms, 
 $ {\bf F}^{(v)}_{\al}$ the force exerted on the same solute atom by the solvent molecules.
Their explicit expressions are given by 
\be
  {\bf F}^{(u)}_{\al}=-\sum_{\beta \neq \al}\frac{\partial U_1(R_{\al \beta})}
{\partial {\bf R}_{\al}}, \qquad
 {\bf F}^{(v)}_{\al} = -\sum_{i=1}^{N} \sum_{a=1}^n 
\frac{\partial U_{int}(|{\bf R}_{\al}-{\bf r}^a_i|)}{\partial {\bf R}_{\al}}
\la{eqn:3.3}
\ee

Our dynamic variable ${\bf A}(t) $ is chosen to be 
\be
  {\bf A}(t)=\left( \begin{array}{r}
					    \Del {\bf R}_{\al} (t)\\ 
						{\bf P}_{\al}(t) \\  
					     \delta\rho^a_{\bf k}(t) \\
						{\bf J}^a_{\bf k}(t)
        			  \end{array} \right)  
\la{eqn:3.4}
\ee
Here $\Del {\bf R}_{\al} (t) $ is the displacement of the position vector ${\bf R}_{\alpha}$ of the $\al$-th solute atom
 from its equilibrium value.
And $ \delta \rho^a_{\bf k}(t)$ is 
 the Fourier component of the density fluctuation 
$\delta \rho^a({\bf r},t)\equiv \rho^a({\bf r},t)-\rho^a_0$ 
($\rho^a_0$ is the average number density of the $a$th atom) of the solvent liquid:
\bea
 \delta \rho^a({\bf r},t) &\equiv & \sum_i \delta 
\big({\bf r}-{\bf r}^a_i(t) \big)-\rho^a_0, \nonumber \\
\delta \rho^a_{\bf k}(t) &=& \int d{\bf r} e^{i{\bf k}\cdot {\bf r}} 
\delta \rho^a({\bf r},t)= \sum_i e^{i{\bf k}\cdot {\bf r}^a_i(t)}-
(2\pi)^3 \rho^a_0 \delta({\bf k})
\la{eqn:3.5}
\eea
Likewise, ${\bf J}^a_{\bf k}(t)$ is the Fourier component of  
the current of the $a$th atom in the solvent liquid:
\bea
\dot  \rho^a_{{\bf k}}(t) &=& 
\sum_i i {\bf k}\cdot \frac{{\bf p}^a_i(t)}{m_a} e^{i{\bf k}\cdot {\bf r}^a_i(t)}
\equiv i{\bf k} \cdot {\bf J}^a_{\bf k}(t) \nonumber \\
{\bf J}^a_{\bf k}(t) &=& 
\sum_i \frac{{\bf p}^a_i(t)}{m_a} e^{i{\bf k}\cdot {\bf r}^a_i(t)} 
\la{eqn:3.6}
\eea
Balucani and Zoppi \cite{balucani} have worked out the special case 
where only the momentum of a solute particle was chosen as a dynamic variable.

We now proceed to obtain the specific expressions of the eqs. (\ref{eqn:2.5})
or (\ref{eqn:2.6}). First one has to compute 
the correlation matrix $\big({\bf A}, {\bf A} \big) $ and its inverse 
$\big({\bf A}, {\bf A} \big)^{-1}$. 
The inner product denotes average over the canonical distribution 
$\exp\big( -\beta {\cal H}(\Gamma) \big)$ with $\beta \equiv 1/(k_B T)$:
\be
\big({\bf a}, {\bf b} \big) \equiv \Big< {\bf a}^* {\bf b} \Big>=
\frac{1}{Z}\int d\Gamma \, {\bf a}^*(\Gamma) {\bf b}(\Gamma) \, 
e^{-\beta {\cal H}(\Gamma)} 
\la{eqn:3.7}
\ee
where $Z$ is the partition function $Z\equiv \int d \Gamma 
\exp\Big(-\beta {\cal H}(\Gamma)\Big)$. \\
 
\subsection {The correlation matrix $\big({\bf A}, {\bf A} \big) $}
The correlation matrix ${\bf C} = \big({\bf A}, {\bf A} \big) $ is given by
\be
 \big({\bf A}, {\bf A} \big) =\left( \begin{array}{rrrr}
	( \Del{\bf R}_{\al}, \Del{\bf R}_{\beta})& ({\bf P}_{\al}, \Del{\bf R}_{\beta})& 
( \delta \rho^a_{\bf k}, \Del{\bf R}_{\beta})& (  {\bf J}^a_{\bf k}, \Del{\bf R}_{\beta})\\ 
		( \Del{\bf R}_{\al}, {\bf P}_{\beta})& ({\bf P}_{\al}, {\bf P}_{\beta})& 
( \delta \rho^a_{\bf k}, {\bf P}_{\beta})& (  {\bf J}^a_{\bf k}, {\bf P}_{\beta})\\
	( \Del{\bf R}_{\al}, \delta \rho^b_{\bf k})& ({\bf P}_{\al}, \delta \rho^b_{\bf k})& 
(\delta \rho^a_{\bf k}, \delta \rho^b_{\bf k})& 
({\bf J}^a_{\bf k}, \delta \rho^b_{\bf k})\\	
	(\Del{\bf R}_{\al}, {\bf J}^b_{\bf k})& ({\bf P}_{\al}, {\bf J}^b_{\bf k})& 
(\delta \rho^a_{\bf k}, {\bf J}^b_{\bf k}  )& 
({\bf J}^a_{\bf k},  {\bf J}^b_{\bf k})		    
        			  \end{array} \right)  
\la{eqn:3.8}
\ee
We first identify the vanishing elements. The following elements vanish:
\bea
(\Del{\bf R}_{\al}, {\bf P}_{\beta})&=& 0, \qquad (\Del{\bf R}_{\al},  {\bf J}^b_{\bf k})=0 \nonumber \\
 ({\bf P}_{\al}, \Del{\bf R}_{\beta})&=& 0, \quad ({\bf P}_{\al}, \delta \rho^b_{\bf k})=0, 
\quad ({\bf P}_{\al},  {\bf J}^b_{\bf k})=0, \nonumber \\
(\delta \rho^a_{\bf k}, {\bf P}_{\beta})&=&0, \quad 
(\delta \rho^a_{\bf k},  {\bf J}^b_{\bf k})=0, 
\nonumber \\
	({\bf J}^a_{\bf k}, \Del{\bf R}_{\beta})&=&0, 
\quad ({\bf J}^a_{\bf k}, {\bf P}_{\beta})=0, 
\quad 
({\bf J}^a_{\bf k},  \delta\rho^b_{\bf k})=0
\la{eqn:3.9}
\eea
They vanish since the momentum integrations 
$$\int d {\bf p}^{nN} \, {\bf p}^a_i \,
\exp \big(-\beta \sum_i \sum_a \frac{ {\bf p}^a_i \cdot {\bf p}^a_i }{2m_a} \big)=0,
\quad \qquad 
\int d {\bf P}^{N_u} \, {\bf P}_{\al} \, \exp \big(-\beta \sum_{\gamma}  
{\bf P}^2_{\gamma}/2M_{\gamma} \big)=0.$$

We now look at the nonvanishing elements.
The momentum correlation of solute particles is easy to compute:
\bea
({\bf P}_{\al}, {\bf P}_{\beta})
&=& 
\frac{1}{Z_P} 
\int d {\bf P}^{N_u} \, {\bf P}_{\al} {\bf P}_{\beta} \,
e^{-\beta \sum_{\gamma} {\bf P}^2_{\gamma}/2M_{\gamma}} \nonumber \\
&=& 
\frac{1}{Z_P} \int d {\bf P}^{N_u} \, {\bf P}_{\al} 
\big(- M_{\beta} k_B T \big) \frac{\partial}{\partial {\bf P}_{\beta}} \,
e^{-\beta \sum_{\gamma} {\bf P}^2_{\gamma}/2M_{\gamma}}
 = k_B T M_{\al} {\bf 1}  \delta_{\al \beta} 
\la{eqn:3.10}
\eea
where $Z_P \equiv \int d {\bf P}^{N_u}  \exp \big(-\beta \sum_{\gamma} {\bf P}^2_{\gamma}/2M_{\gamma} \big)$, 
and ${\bf 1}$ is the unit $(3\times3)$ matrix.
The eq. (\ref{eqn:3.10}) is nothing but the equipartition theorem.

Since the  general current-current correlation function
 $({\bf J}^a_{\bf k},{\bf J}^b_{\bf k}) $ will  
 have non-vanishing correlation between the same Cartesian components only, 
 it is sufficient to define the current-current correlation function as
\be
J_{ab}(k) \equiv \frac{1}{N} \Big< {\bf J}^a_{-\bf k} \cdot {\bf J}^b_{\bf k}  \Big>
\la{eqn:3.11}
\ee
Its calculation is somewhat involved:
\bea
J_{ab}(k)
&=& \frac{1}{N}
\sum_i \sum_j \frac{1}{m_a}\frac{1}{m_b} 
\Big< {\bf p}^a_i \cdot {\bf p}^b_j e^{-i{\bf k}\cdot \big( {\bf r}^a_i - {\bf r}^b_j \big)}  \Big>  \nonumber \\
&=& \frac{1}{N}
\sum_i \sum_j \frac{1}{m_a}\frac{1}{m_b} 
\Big< {\bf p}^a_i \cdot {\bf p}^b_j \Big> \Big< e^{-i{\bf k}\cdot \big( {\bf r}^a_i - {\bf r}^b_j \big)}  \Big>  \nonumber \\
&=& \frac{1}{N}
\sum_i \sum_j 
\Big< {\bf v}^a_i  \cdot {\bf v}^b_i \Big> \delta_{ij} \Big< e^{-i{\bf k}\cdot \big( {\bf r}^a_i - {\bf r}^b_j \big)}  \Big>  \nonumber \\
&=& 
\frac{1}{N}
\sum_i 
\Big< {\bf v}^a_i  \cdot {\bf v}^b_i \Big>  \Big< e^{-i{\bf k}\cdot \big( {\bf r}^a_i - {\bf r}^b_i \big)}  \Big>
\la{eqn:3.12}
\eea
A general expression of this quantity is given in Eq. (7) in \cite{yama-chong-hirata}. 

The remaining elements  $( \Del{\bf R}_{\al}, \Del{\bf R}_{\beta})$, 
$( \Del{\bf R}_{\al}, \delta \rho^b_{\bf k})$, $(\delta \rho^a_{\bf k}, \Del{\bf R}_{\beta})$, 
and $(\delta \rho^a_{\bf k}, \delta \rho^b_{\bf k})$  involve
the spatial coordinates only. 
We consider them in order. 
In the present work we will not specify particular form for correlation of 
initial position of solute particles since here we are interested in laying out general structure of the dynamics.
We first have the displacement correlation matrix for the solute particles
\be 
{\bf L}_{\al \beta} \equiv \Big( \Del{\bf R}_{\al}, \, \Del{\bf R}_{\beta} \Big)  
\la{eqn:3.13}
\ee
where ${\bf L}_{\al \beta}$ is a $(3N_u\times3N_u)$ matrix.
We next consider $(\Del{\bf R}_{\al}, \delta \rho^b_{\bf k})$.
First note that $(\Del{\bf R}_{\al}, \delta \rho^b_{\bf k})
=\big<   \Del{\bf R}_{\al} \rho^b_{{\bf k}}\big>-(2\pi)^3 \rho^b_0 \delta({\bf k}) 
\big<\Del{\bf R}_{\al} \big>
= \big< \Del{\bf R}_{\al} \rho^b_{\bf k} \big>$ since $\big<\Del{\bf R}_{\al} \big>=0$.
Therefore
\be
{\bf B}^{\al, b}_{\bf k}\equiv (\Del{\bf R}_{\al}, \delta \rho^b_{\bf k})
= \big< \Del{\bf R}_{\al} \, \rho^b_{\bf k} \big>
\la{eqn:3.14}
\ee
In Appendix A, we show that this quantity and its transposed one vanish;
\be
 {\bf B}_{\al,b}({\bf k})={\bf B}_{a,\beta}({\bf k})={\bf 0}
\la{eqn:3.15}
\ee

Finally, we have the static structure factor of solvent molecular liquid defined as
\bea
\chi_{ab} ({\bf k}) &\equiv& \frac{1}{N}(\delta \rho^a_{\bf k},  \delta \rho^b_{\bf k})
= \frac{1}{N}\Big< \delta \rho^a_{-\bf k}  \delta \rho^b_{\bf k}\Big>
\la{eqn:3.16}
\eea
This can be calculated using the RISM theory.

Summing up the above results, we have the following block-diagonal matrix 
for $({\bf A}, {\bf A})$.
\be
 \big({\bf A}, {\bf A} \big) =\left( \begin{array}{rrrr}
	 {\bf L}_{\al \beta} & {\bf O} & \quad {\bf 0}& {\bf 0}\\ 
		{\bf O}& \quad k_B T M_{\al}  {\bf 1} \delta_{\al \beta}  &{\bf 0}& {\bf 0}\\
	{\bf 0}^T& {\bf 0}^T& \quad \qquad N\chi_{ab}({\bf k})&{\bf \u{0}} \\
	{\bf 0}^T& {\bf 0}^T& {\bf \u{0}}^T & \qquad N J_{ab}(k)  
        			  \end{array} \right)  
\la{eqn:3.17}
\ee
where ${\bf O}$ denotes the $(3N_u\times3N_u)$ zero matrix, ${\bf 0}$  
the $(3N_u \times n)$ zero matrix, ${\bf \u{0}}$ the $(n \times n)$ zero matrix, and the superscript $T$  the transpose
matrix.

\subsection{Inverse of $({\bf A}, {\bf A})$} 
Since the above correlation matrix is block-diagonal, it is trivial to 
obtain the inverse $({\bf A}, {\bf A})^{-1}$ as
\be
 \big({\bf A}, {\bf A} \big)^{-1} =\left( \begin{array}{rrrr}
\big({\bf L}^{-1}\big)_{\al \beta} & 
 {\bf O} &  {\bf 0}& {\bf 0}\\ 
		{\bf O}& \qquad \frac{1}{k_BTM_{\al}} {\bf 1} \delta_{\al \beta}&{\bf 0}& {\bf 0}\\
	{\bf 0}^T &{\bf 0}^T&\qquad \frac{1}{ N} \chi^{-1}_{ab}(k) &{\bf \u{0}} \\
	{\bf 0}^T& {\bf 0}^T& {\bf \u{0}}&\qquad \frac{1}{N} J^{-1}_{ab}(k)
        			  \end{array} \right)  
\la{eqn:3.18}
\ee
Here the inverse matrices $\big({\bf L}^{-1}\big)_{\al \beta}$, $ \chi^{-1}_{ab}(k)$ and $J^{-1}_{ab}(k) $ are defined as
$$
\sum_{\gamma=1}^{N_u} {\bf L}_{\al \gamma} \big({\bf L}^{-1}\big)_{\gamma \beta} = {\bf 1} \delta_{\al \beta}, \qquad
\sum_{c=1}^n \chi_{ac}(k) \chi^{-1}_{cb}(k) = \delta_{ab}, \qquad
\sum_{c=1}^n J_{ac}(k) J^{-1}_{cb}(k) = \delta_{ab}.
$$

\subsection{The frequency matrix $i{\bf {\Omega}}$}
Here we compute the frequency matrix $i{\bf {\Omega}}$ which is defined as
\be
i{\bf {\Omega}}_{\lambda \nu} \equiv \sum_{\lambda'} 
({\bf A}_{\lambda'}, \dot{\bf A}_{\lambda})
 \big[({\bf A}, {\bf A})^{-1}\big]_{\lambda' \nu}
 \la{eqn:3.19}
\ee
We first look at the elements of the matrix $({\bf A}, \dot{\bf A}) $:
\be
 ({\bf A}, \dot{\bf A}) =\left( \begin{array}{rrrr}
	( \Del{\bf R}_{\al}, \Del\dot{\bf R}_{\beta})&({\bf P}_{\al}, \Del\dot{\bf R}_{\beta}) & 
(\delta \rho^a_{\bf k}, \Del\dot{\bf R}_{\beta}) & ({\bf J}^a_{\bf k}, \Del\dot{\bf R}_{\beta})\\ 
	(\Del{\bf R}_{\al}, \dot{\bf P}_{\beta})	& ({\bf P}_{\al}, \dot{\bf P}_{\beta})& (\delta \rho^a_{\bf k}, \dot{\bf P}_{\beta})
& ({\bf J}^a_{\bf k}, \dot{\bf P}_{\beta})\\
	(\Del{\bf R}_{\al}, \dot\rho^b_{\bf k})& ({\bf P}_{\al}, \dot \rho^b_{\bf k})
& 
(\delta \rho^a_{\bf k}, \dot \rho^b_{\bf k})& ({\bf J}^a_{\bf k},  \dot \rho^b_{\bf k})
\\	
	(\Del{\bf R}_{\al},  \dot {\bf J}^b_{\bf k})&({\bf P}_{\al},  \dot {\bf J}^b_{\bf k})
 & (\delta \rho^a_{\bf k},  \dot {\bf J}^b_{\bf k})
& 
({\bf J}^a_{\bf k},  \dot {\bf J}^b_{\bf k})		    
        			  \end{array} \right)  
\la{eqn:3.20}
\ee
First we obtain some elements of $\dot{\bf A}$ using the Liouville operator
(\ref{eqn:3.2}).
\bea
\Del\dot {\bf R}_{\al} &=& i{\cal L} \Del{\bf R}_{\al}=\frac{{\bf P}_{\al}}{M_{\al}} \nonumber \\
\dot {\bf P}_{\al}&=& i{\cal L} {\bf P}_{\al} ={\bf F}_{\al}  \nonumber \\
\dot \rho^a_{\bf k}&=& i{\cal L}\rho^a_{\bf k}=i{\bf k}\cdot {\bf J}^a_{\bf k} 
\nonumber \\
\dot {\bf J}^a_{\bf k}  &=& 
\sum_i \frac{1}{m_a} \Big(  \dot {\bf p}^a_i + {\bf p}^a_i 
i{\bf k} \cdot \frac{ {\bf p}^a_i}{m_a}  \Big) e^{i{\bf k}\cdot {\bf r}^a_i}
\la{eqn:3.21}
\eea
where ${\bf F}_{\al}$ is the total force exerted on the $\al$th solute particle 
by the solvent as well as by other solute particles.
Actually when we compute the elements involving 
$\dot {\bf P}$ or $\dot {\bf J}^a_{\bf k}$, 
it is more convenient to use the integration by parts.
It is useful to remember that whereas $\Del\dot {\bf R}$ and $\dot \rho^a_{\bf k}$
involve single momentum (${\bf P}$ or ${\bf p}_i$), 
$\dot {\bf P}$ and $\dot {\bf J}^a_{\bf k}$
involve zero (since ${\bf p}^a_i$ is the force acting on the $a$th atom of the 
$i$th molecule, which only involves the positions of solute particles and 
solvent molecules), or two momentums (two ${\bf p}_i$).
Using this fact, we can easily identify the vanishing elements:
\bea
(\Del{\bf R}_{\al}, \Del\dot{\bf R}_{\beta})&=& 0, \quad (\Del{\bf R}_{\al},  \dot \rho^b_{\bf k})=0, 
 \quad (\Del{\bf R}_{\al},  \dot {\bf J}^b_{\bf k})=0 \nonumber \\
 ({\bf P}_{\al}, \dot{\bf P}_{\beta})&=& 0, \quad ({\bf P}_{\al}, \dot \rho^a_{\bf k})=0, 
\quad ({\bf P}_{\al},  \dot {\bf J}^b_{\bf k})=0, \nonumber \\
(\delta \rho^a_{\bf k}, \Del\dot{\bf R}_{\beta})&=&0, \quad 
(\delta \rho^a_{\bf k}, \dot{\bf P}_{\beta})=0, 
\quad (\delta \rho^a_{\bf k}, \dot \rho^b_{\bf k})=0, \nonumber \\
({\bf J}^a_{\bf k}, \Del\dot{\bf R}_{\beta})&=&0, \quad ({\bf J}^a_{\bf k}, \dot{\bf P}_{\beta})=0, \quad 
({\bf J}^a_{\bf k},  \dot {\bf J}^b_{\bf k})=0
\la{eqn:3.22}
\eea
The nonvanishing elements are 
\bea
(\Del{\bf R}_{\al},\dot{\bf P}_{\beta})&=& -\frac{1}{M}({\bf P}_{\al},{\bf P}_{\beta})=-k_BT {\bf 1} \delta_{\al \beta}\nonumber \\
({\bf P}_{\al},\Del\dot{\bf R}_{\beta})&=& \frac{1}{M}({\bf P}_{\al},{\bf P}_{\beta})=k_BT {\bf 1} \delta_{\al \beta} \nonumber \\
(\delta \rho^a_{\bf k},  \dot {\bf J}^b_{\bf k})
&=& 
(\dot \delta \rho^a_{\bf k},  {\bf J}^b_{\bf k}) =
i{\bf k} \cdot ({\bf J}^a_{\bf k},  \dot {\bf J}^b_{\bf k}) 
=iN{\bf k} J_{ab}(k)  \nonumber \\
( {\bf J}^a_{\bf k}, \dot \rho^b_{\bf k}) &=& 
(  {\bf J}^a_{\bf k}, {\bf J}^b_{\bf k}) \cdot i{\bf k} 
=iN{\bf k} J_{ab}(k)
\la{eqn:3.23}
\eea
Taking all these into account, we obtain
\be
 i{\bf {\Omega}} =\left( \begin{array}{rrrr}
	{\bf O} &k_BT{\bf 1}\delta_{\al \beta} & {\bf 0}& {\bf 0}\\ 
		-k_BT{\bf 1}\delta_{\al \beta}& \quad {\bf O}&{\bf 0}& {\bf 0}\\
	{\bf 0}^T& \quad{\bf 0}^T& {\bf \u{0}}& \quad iN{\bf k} J_{ab}(k)\\
	{\bf 0}^T& \quad{\bf 0}^T&\quad iN{\bf k} J_{ab}(k) &{\bf \u{0}} 
        			  \end{array} \right)  \cdot  ({\bf A}, {\bf A})^{-1}
\la{eqn:3.24}
\ee
Using the inverse correlation matrix
(\ref{eqn:3.18}), we compute $i{\bf {\Omega}}$ as
\be
i{\bf {\Omega}} =
\left( \begin{array}{rrrr}
	{\bf O} & \frac{1}{M_{\al}}{\bf 1} \delta_{\al \beta}& {\bf 0}& {\bf 0}\\ 
		-k_BT \big({\bf L}^{-1}\big)_{\al \beta} & {\bf O}& {\bf 0} & {\bf 0}\\
	{\bf 0}^T & {\bf 0}^T& {\bf \u{0}}& \qquad i{\bf k} \delta_{ab}\\
	{\bf 0}^T & \quad{\bf 0}^T&\qquad i{\bf k} 
\sum_{c=1}^n J_{ac}(k)\chi^{-1}_{cb}(k) &{\bf \u{0}} 
        			  \end{array} \right) \nonumber \\
 \la{eqn:3.25}       			  
\ee

\subsection{The reversible part}
From (\ref{eqn:2.5}), the reversible part of the Langevin equation is given by
$i{\bf \Omega}\cdot {\bf A}(t)$. Using (\ref{eqn:3.25}), we obtain
\be
i{\bf {\Omega}} \cdot {\bf A}(t)=\left( \begin{array}{r}
	 {\bf P}_{\al}(t)/M_{\al}\\ 
		-k_BT \sum_{\beta} \big({\bf L}^{-1}\big)_{\al \beta} \cdot \Del{\bf R}_{\beta}(t)\\
 i{\bf k} \cdot {\bf J}^a_{{\bf k}}(t) \\
	i{\bf k} 
\sum_{b,c} J_{ac}(k)\chi^{-1}_{cb}(k)\delta \rho^b_{{\bf k}}(t) 
        		\end{array} \right) 
 \la{eqn:3.26}       			  
\ee

\subsection{The fluctuating force}
The fluctuating force at $t=0$ from (\ref{eqn:2.6}) is given by 
\be
{\bf f}=(1-{\cal P}) \dot{\bf A}
=\dot{\bf A}-i{\bf {\Omega}}\cdot {\bf A}
 \la{eqn:3.27}
\ee
where we used 
${\cal P} \dot{\bf A}=({\bf A}, \dot{\bf A})\cdot ({\bf A}, {\bf A})^{-1}{\bf A}
=i{\bf {\Omega}}\cdot {\bf A}$.
We first obtain 
\be
 \dot {\bf A}=\left( \begin{array}{r}
	 \Del\dot {\bf R}_{\al}\\ 		\dot {\bf P}_{\al}\\
 \dot \rho^a_{{\bf k}} \\
	\dot {\bf J}^a_{{\bf k}} 
        			  \end{array} \right) 
        			  =\left( \begin{array}{r}
	 {\bf P}_{\al}/M_{\al}\\ 		{\bf F}_{\al}\\
 i{\bf k}\cdot {\bf J}^a_{{\bf k}} \\
	\dot {\bf J}^a_{{\bf k}} 
        			  \end{array} \right)
	 \la{eqn:3.28}       			  
\ee
and 
\be
i{\bf {\Omega}} \cdot {\bf A}=\left( \begin{array}{r}
	 {\bf P}_{\al}/M_{\al}\\ 
		-k_BT \sum_{\beta} \big({\bf L}^{-1}\big)_{\al \beta} \cdot \Del{\bf R}_{\beta}\\
 i{\bf k} \cdot {\bf J}^a_{{\bf k}} \\
i{\bf k} 
\sum_{b,c} J_{ac}(k)\chi^{-1}_{cb}(k)\delta \rho^b_{{\bf k}}  
        		\end{array} \right) 
 \la{eqn:3.29}       			  
\ee
which is obtained by setting $t=0$ in (\ref{eqn:3.26}).
Using the above two results, we obtain for the fluctuating force as
\be
 {\bf f}=\left( \begin{array}{r}
	 {\bf 0}\\ 
	{\bf  W}_{\al}	\\
 0\\
	 {\bf \Xi}^a_{{\bf k}}
        			  \end{array} \right), 
        			  \qquad
        			  {\bf f}(t)=e^{i t (1-{\cal P})  {\cal L}} \left( \begin{array}{r}
	 {\bf 0}\\ 
	{\bf  W}_{\al}\\
 0\\
	 {\bf \Xi}^a_{{\bf k}}
        			  \end{array} \right)
 \la{eqn:3.30}       			  
\ee
where 
\be
{\bf W}_{\al}  \equiv {\bf F}_{\al}+ k_BT \sum_{\beta} \big({\bf L}^{-1}\big)_{\al \beta} \cdot \Del{\bf R}_{\beta}, \qquad
{\bf \Xi}^a_{{\bf k}} 
\equiv \dot {\bf J}^a_{{\bf k}}-i{\bf k} 
\sum_{b,c} J_{ac}(k)\chi^{-1}_{cb}(k)\delta \rho^b_{{\bf k}} 
\la{eqn:3.31} 
\ee

\subsection{The memory matrix}
The memory function matrix ${\bf K}(t)$ is calculated as
\bea
{\bf K}(t)&\equiv&
 \big({\bf f}, {\bf f}(t)\big) \big[ \big({\bf A}, {\bf A}\big)^{-1}\big]\nonumber \\
&=& 
\left( \begin{array}{rrrr}
	{\bf O} & {\bf O} & {\bf 0}& {\bf 0}\\ 
		{\bf O} & \frac{1}{k_BTM_{\al}}({\bf W}_{\al}, {\bf W}_{\beta}(t)) &
{\bf 0}& \frac{1}{N} \sum_b J^{-1}_{ab}(k)({\bf \Xi}^b_{\bf k}, {\bf W}_{\beta}(t))\\
	{\bf 0}^T & {\bf 0}^T & {\bf 0}& {\bf 0}\\
	{\bf 0}^T &\frac{1}{Mk_BT}({\bf W}_{\al}, {\bf \Xi}^b_{\bf k}(t))&{\bf 0}&
 \frac{1}{N} \sum_c J^{-1}_{ac}(k) ({\bf \Xi}^c_{\bf k}, {\bf \Xi}^b_{\bf k}(t))  
        			  \end{array} \right) 
   \la{eqn:3.32} 
\eea
where $ {\bf W}(t)\equiv \exp \Big( it  (1-{\cal P}) {\cal L} \Big) {\bf W}$ 
and $ {\bf \Xi}^a_{{\bf k}}(t)\equiv \exp \Big( it (1-{\cal P}) {\cal L} \Big) 
{\bf \Xi}^a_{{\bf k}}$. 

In (\ref{eqn:3.32}), the two terms exhibit explicit $N$-dependence.
In the thermodynamic limit in which $N$ is taken to be infinite while
the number of solute particles $N_u$ remains finite, 
the term $\frac{1}{N} \sum_b J^{-1}_{ab}(k)({\bf \Xi}^b_{\bf k}, {\bf W}(t)) $ will vanish since the ensemble average $\big({\bf \Xi}_{\bf k}, {\bf W}(t) \big) $ will remain finite.
The other term $\frac{1}{N} \sum_c J^{-1}_{ac}(k) 
({\bf \Xi}^c_{\bf k}, {\bf \Xi}^b_{\bf k}(t))  $ will not vanish since  the ensemble average $\big({\bf \Xi}^c_{\bf k}, {\bf \Xi}^b_{\bf k}(t)\big)$  is proportional to $N$.  
Therefore only the latter term survives in the thermodynamic limit. 

In Appendix B, we show that 
\be
({\bf W}_{\al}, {\bf \Xi}^b_{\bf k}(t))=0
\la{eqn:3.33}
\ee
Therefore the final expression of the memory matrix is given by
\be
{\bf K}(t)=
\left( \begin{array}{rrrr}
	{\bf O} & {\bf O} & {\bf 0}& {\bf 0}\\ 
		{\bf O} & \frac{1}{k_BTM_{\al}}({\bf W}_{\al}, {\bf W}_{\beta}(t)) &
{\bf 0}& {\bf 0}\\
	{\bf 0}^T & {\bf 0}^T &  0& 0\\
	{\bf 0}^T &{\bf 0}^T&0& \frac{1}{N} \sum_c J^{-1}_{ac}(k) ({\bf \Xi}^c_{\bf k},
 {\bf \Xi}^b_{\bf k}(t))  
        			  \end{array} \right) 
   \la{eqn:3.34} 
\ee

\subsection{The explicit form of the exact dynamic equations}
With the explicit results of the previous sections, 
we here write down an exlicit form for the time evolution equation (\ref{eqn:2.5})
\bea
\frac{d \Del{\bf R}_{\al}(t)}{dt}&=& \frac{{\bf P}_{\al}(t)}{M_{\al}},  \nonumber \\
&{}& \nonumber \\
\frac{d{\bf P}_{\al}(t)}{dt}&=& -k_B T \sum_{\beta} \big({\bf L}^{-1}\big)_{\al \beta} \cdot \Del{\bf R}_{\beta}(t)
-\int_0^t ds \, \sum_{\beta}
{\bf \Gamma}_{\al \beta}(t-s) \cdot \frac{{\bf P}_{\beta}(s)}{M_{\beta}}
+{\bf W}_{\al}(t), \nonumber \\
&{}& \nonumber \\
\frac{d \delta \rho^a_{\bf k}(t)}{dt}&=& i{\bf k} \cdot {\bf J}^a_{\bf k}(t),  \nonumber \\
&{}& \nonumber \\
\frac{d {\bf J}^a_{\bf k}(t)}{dt} &=& 
i{\bf k} 
\sum_{b,c} J_{ac}(k)\chi^{-1}_{cb}(k)\delta \rho^b_{{\bf k}}(t) 
- \frac{1}{N} \sum_{b,c} J^{-1}_{ac}(k) \int_0^t ds \, 
 {\bf M}^{bc}_{\bf k} (t-s) \cdot {\bf J}^b_{\bf k}(s)
 +  {\bf \Xi}^a_{\bf k}(t)   \nonumber \\
 \la{eqn:3.35}
 \eea
  In the above set of dynamic equations for the solute and solvent molecules, 
the random forces take the following forms
 \bea
  {\bf W}_{\al} (t) &=& e^{it (1-{\cal P}) {\cal L}} \Big( {\bf F}_{\al}+ k_BT \sum_{\beta} \big( {\bf L}^{-1} \big)_{\al \beta} \cdot \Del{\bf R}_{\beta} \Big), \nonumber \\
 {\bf \Xi}^a_{\bf k}(t) &=& e^{it (1-{\cal P}){\cal L}} 
 \Big( 
 \dot {\bf J}^a_{{\bf k}}-i{\bf k} 
\sum_{b,c} J_{ac}(k)\chi^{-1}_{cb}(k)\delta \rho^b_{{\bf k}} \Big)
\la{eqn:3.36}
\eea
The memory functions in (\ref{eqn:3.35}) are given by 
the time correlations of the random forces;
\be
{\bf \Gamma}_{\al \beta}(t) =\frac{1}{k_B T} \Big< {\bf W}_{\al}(t) {\bf W}_{\beta}(0)  \Big>, 
\qquad   \qquad
{\bf M}^{bc}_{\bf k} (t) =\Big< {\bf \Xi}^b_{\bf k}(t) \, {\bf \Xi}^c_{-\bf k}(0) \Big>.
\la{eqn:3.37}
\ee

\section{Discussions}
\setcounter{equation}{0}
\subsection{Solvent dynamics}
When the fluid is far from protein, or in bulk, where perturbation from protein vanishes, the last two expressions concerning solvent in Eq. (III.35) reduce to the equations for pure-water dynamics, Eqs. (24) and (25), derived by one of the authors 
\cite{hirata92}, except for an approximation made in the factor $Jab$ \cite{jab}.  
The equation is further simplified to produce the site-site Smoluchowski-Vlasov (SSSV) equation 
if one makes the memory function local in time as well as in space. The equation can be analytically solved by means of the Laplace transform to produce the van Hove or space-time correlation function of water, with the {\em input} of the site-site pair correlation functions of the solvent  obtained from the RISM theory. The theory has been successfully applied to a variety of solvent relaxation processes induced by an abrupt change in the electronic structure of a solute molecule, or solvation dynamics, which can be probed by 
the dynamic Stokes-shift \cite{stokes_shift}. 
    So, the two equations concerning solvent in (III.35) can be regarded as a generalization of the previous theories developed for pure water to that subject to the field exerted from protein atoms. 
There are several remarks to be made with respect to the generalization. 
Firstly, the translational invariance of the system is no longer valid. 
Therefore, the equations should be solved in three-dimensional Cartesian-space. 
Secondly, the factor 
$\chi_{ab}({\bf r}, {\bf r}')=N^{-1}\big< \delta\rho^a({\bf r}) \delta\rho^b({\bf r}')\big>$  appearing in the equation is a two body density correlation function, but subject to the "external force" due to protein. Such a theory for obtaining the function is under development, but it is too primitive at the moment to be applied to the problem we are facing. Therefore, we may adopt the superposition approximation $\big<\delta\rho^a({\bf r})\delta\rho^b({\bf r}')\big> = \big<\delta\rho^a({\bf r})\big> 
\big<\delta\rho^b({\bf r}') \big>$  to this case. 
Then, $\big<\delta\rho^a({\bf r})\big>$ can be readily evaluated from the 3D-RISM theory. 
 
A number of possible applications of the dynamic equations for the solvent are conceivable. 
An interesting example is the current-current correlation function 
$\big<{\bf J}^a({\bf r},0){\bf J}^b({\bf r}',t)>$ of water and ions in a molecular channel, which is concerned with 
many observables including the permeability of water and ions across the cell membrane \cite{phong}.
The equation for the correlation function can be readily obtained by coupling the two equations for solvent
with the aid of the 3D-RISM/RISM theory.

\subsection{Solute dynamics}
The first two equations in Eq. (III.35) concerning solute dynamics are combined together to result in
\be
M_{\al} \frac{d^2 \Del{\bf R}_{\al}(t)}{dt^2}+k_BT \sum_{\beta} \big({\bf L}^{-1}\big)_{\al \beta} \cdot \Del{\bf R}_{\beta}(t)
+\int_0^t ds \, \sum_{\beta} {\bf \Gamma}_{\al \beta}(t-s) \cdot 
\frac{d \Del{\bf R}_{\beta}(s)}{ds}
={\bf W}_{\al}(t)
\la{eq4.1}
\ee
The equation is regarded as a generalization of the equation for a coupled set of  Langevin-oscillators, 
first examined by Wang and Uhlenbeck \cite{wang}, to a realistic model of protein in water. 
Wang and Uhlenbeck proposed a model in which a coupled set of oscillators consisting of spherical beads is immersed in a viscous liquid, 
and applied the Langevin theory to the oscillators. 
Later on, Lamm and Szabo \cite{lamm} performed a normal mode analysis on the Wang-Uhlenbeck oscillators, assuming a phenomenological friction term. 
Kottlam and Case \cite{case}, and Ansari \cite{ansari} applied the Langevin mode method of Lamm and Szabo to proteins. The same method was also applied to the dynamics of 
DNA \cite{shih} and RNA \cite{zacharias} in solvents.
A review on the normal mode analysis in general
(including the Langevin mode analysis) in the dynamics of biomolecules
is presented in \cite{jma}.

There are several comments to be made on the new equation (\ref{eq4.1}). 
Firstly, the equation does not include a term related to the force which originates from the first derivative of the free energy surface with respect to the position. The force acting on an atom of protein comes from the three contributions, one which is proportional to the displacement of the atom from its equilibrium position (the second term in the left hand side in (\ref{eq4.1})), and the friction term proportional to the velocities of 
the atoms, and that due to the random force (the term in the right hand side in (\ref{eq4.1})). 
The physical origin as to why the equation does not include the first derivative of the free energy lies in our treatment based on the generalized Langevin theory. The whole idea of the generalized Langevin theory is to project all the degrees of freedom in the phase space onto few dynamic variables under concern. The projection is carried out using a projection operator, defined by Eqs. (II.3) and (II.4), in terms of an ensemble average of two variables which are fluctuating around an equilibrium in the phase space.  
Obviously by definition, the ensemble average of the displacement of atoms in protein should be zero in equilibrium.
             
Such a force as the first derivative of the free energy, which may cause the complete shift of the equilibrium, is not included in the treatment. The situation is somewhat analogous to the case of a harmonic oscillator, in which an oscillator swings back and force around a minimum of the harmonic potential. Only force acts on the system is the restoring force proportional to the displacement from the potential minimum. In our case, too, only force acting on the protein atoms is the one which restores atom positions from fluctuating to equilibrium ones. However, there is an essential difference in physics between the two systems. The equilibrium position of a harmonic oscillator is the minimum of mechanical potential energy, while that of protein in water is the minimum in the thermodynamic potential or the free energy, which is concerned not only with energy but also with the entropy both of protein and of water. So, in the case of protein in water, the stochastic character of the dynamics is attributed not only to the random force term, but also to the conformational fluctuation of protein around its equilibrium state, induced by solvent, while the stochastic character is resulted just from the random force term in the case of the coupled harmonic oscillators treated by Wang and Unlenbeck.

The argument above suggests interesting physics implied in Eq. (\ref{eq4.1}), and its application to biological functions. 
If one ignores the friction and random force terms in Eq. (\ref{eq4.1}),  one gets
\be
 M_{\al} \frac{d^2 \Del{\bf R}_{\al}(t)}{dt^2}
=-k_BT \sum_{\beta} \big({\bf L}^{-1}\big)_{\al\beta} \cdot \Del{\bf R}_{\beta}(t)
\la{eq4.2}
\ee     
This equation can be viewed as a coupled set of "harmonic oscillators", whose "Hessian" matrix is given by
$k_B T \big({\bf L}^{-1}\big)_{\al \beta} $. 
Considering Eq. (III.13), the "Hessian" matrix is related to the variance-covariance matrix of the positional fluctuation by 
 \be
 k_BT {\bf L}^{-1} \equiv k_B T  \Big<  \Delta {\bf R} \,\, \Delta {\bf R}   \Big>^{-1}
 \la{eq4.3}
\ee  
 The observation strongly suggests that the dynamics described by Eq. (\ref{eq4.2}) 
 is that of fluctuation around a minimum of the free energy surface consisting not only of the interactions among atoms in the protein, but of the solvation free energy. In this respect, the configuration corresponding to the free energy minimum is not just one but an ensemble of distinguishable configurations concerning protein and solvent, which can be converted among each other due to the thermal noise. The free energy surface can be 
given by
\be
F \big(\{\Del{\bf R}\} \big) = U\big(\{\Del{\bf R}\} \big) + \Delta \mu\big(\{\Del{\bf R}\} \big)
 \la{eq4.4}
\ee  
 where $U\big(\{\Del{\bf R}\} \big) $ is the interaction potential energy among atoms in a protein, and 
$\Delta \mu\big(\{\Del{\bf R}\} \big)$ is the solvation free energy of protein whose
 conformation  is $\{{\bf R}\}$ \cite{kinoshita}.

The above consideration further suggests a method to evaluate the variance-covariance matrix, which characterizes structural fluctuation of protein, based on the 3D-RISM theory. 
The variance-covariance matrix is closely related to the Hessian matrix, Eq. (\ref{eq4.3}), and the Hessian matrix is the second derivative of the free energy surface, namely,
  \be
k_B T \big({\bf L}^{-1}\big)_{\al \beta} = 
\frac{\partial^2 F \big(\{\Del{\bf R}\} \big)}
{\partial \Del{\bf R}_{\al} \partial \Del{\bf R}_{\beta}}
\la{eq4.5}
\ee    
Since the free energy $F \big(\{\Del{\bf R}\} \big)$ can be obtained by solving the 3D-RISM/RISM equation, Eq.(\ref{eq4.5}) provides a way to evaluate 
the variance-covariance matrix. 

The variance-convariance matrix is by itself quite informative for characterizing the structural fluctuation of protein around its native state in atomic detail. As an example, let us consider a hinge-bending motion of protein. The variance-covariance matrix should have a structure in which a block of elements $\big<\Delta {\bf R}_{\al} \Delta {\bf R}_{\beta} \big>$ for atom pairs, $\al, \beta$, belonging to the two sides of the hinge-axis, have the negative sign because the direction of the displacements 
$ \Delta {\bf R}_{\al}$ and $\Delta {\bf R}_{\beta} $ is opposite. 

Usefulness of the variance-covariance matrix is not limited to characterization of the structural fluctuation around an equilibrium state. 
The Eq. (\ref{eq4.5}) implies that the free energy of protein at an equilibrium conformation takes the form
\be
F \big(\{\Del{\bf R}\} \big) 
=\frac{1}{2} k_B T \sum_{\al,\beta} 
\Del{\bf R}_{\al} \cdot \big({\bf L}^{-1}\big)_{\al \beta} \cdot \Del{\bf R}_{\beta}.
\la{eq4.6}
\ee
In the presence of a small perturbation due to, say, ligand binding, 
the above free energy can be changed due to the perturbation as
\be
F \big(\{\Del{\bf R}\} \big) 
=\frac{1}{2} k_B T \sum_{\al,\beta} 
\Del{\bf R}_{\al} \cdot \big({\bf L}^{-1}\big)_{\al \beta} \cdot \Del{\bf R}_{\beta}
-\sum_{\al} \Del{\bf R}_{\al} \cdot {\bf f}_{\al}
\la{eq4.7}
\ee
where ${\bf f}_{\al}$ is the force acting on the $\al$th protein atom due to
the perturbation.
Then, the conformational change due to the perturbation can be determined by the variational principle 
\be
\frac{\partial F}{\partial \Del{\bf R}_{\al}} =0.
\la{eq4.8}
\ee
With Eq. (\ref{eq4.7}), Eq. (\ref{eq4.8}) gives 
\be
\Big<   \Delta {\bf R}_{\al} \Big>_1  = \big( k_B T \big)^{-1}
 \sum_{\beta}   \Big<   \Delta {\bf R}_{\al} \Delta {\bf R}_{\beta} \Big>_0 
 \cdot {\bf f}_{\beta} 
\la{eq4.9}
\ee  
where the subscript $1(0)$ denotes the presence (absence) of the perturbation. 
Therefore, Eq. (\ref{eq4.5}) combined with Eq.(\ref{eq4.9}) provides 
a theoretical basis for analyzing the conformational relaxation of protein in water due to a perturbation such as ligand binding. Th Eq.(\ref{eq4.9}) is first derived by 
Ikeguchi et. al. \cite{ikeguchi} based on the linear response theory.  

The equation (\ref{eq4.1}) is also a generalized equation which provides molecular basis for the phenomenological Rouse-Zimm model of the polymer dynamics \cite{doi}, with a proper account of the variance-covariance matrix, the diagonal terms of which correspond to the mean square displacement of each atom in equilibrium states. This suggests that the theory can be applied not only to the native conformation of protein but also to characterizing the denatured or random-coil state. However, the application requires special care of the ensemble average to evaluate the variance-covariance matrix, since the average, by definition, should be taken over virtually an infinite number of conformations randomly appearing in the solution. Nevertheless, a practical method to evaluate the variance- covariance matrix for the random-coil state of protein can be suggested based on the 3D-RISM/RISM theory as follows. First, produce some small number of conformations for protein in water by means of a generalized ensemble technique such as the replica-exchange algorithm. Second, evaluate the second derivative of the free energy surface of each conformation based on Eq. (\ref{eq4.5}), and take the average of the results over the conformations, which will give rise to the variance-covariance matrix for the sampled conformational space. Third, add more conformations to the sample to take the average. Repeat the procedure until the convergence is attained. Our implication is that the convergence will be attained rather quickly, because the variance-covariance matrix for each conformation, obtained from Eq. (\ref{eq4.5}), is already an average over a large number of conformations in the free energy surface. The converged variance-covariance matrix can be compared with observable quantities which characterize a random coil state of protein, such as the gyration radius and the distribution  of end-to-end distance.

\section{Concluding Remarks}
In the present work, we have proposed a new theory of dynamics based on the generalized Langevin theory, which can be applied to structural fluctuation around a native state of protein in water. The displacement vector of atom positions and their conjugated momentum, are chosen for dynamic variables for protein, while the density fields of atoms and their momentum fields are chosen for water. Projection of other degrees of freedom onto those dynamic variables using the standard projection operator method produced essentially two equations which describe the time evolution of fluctuation concerning the density field of solvent and the conformation of protein around an equilibrium state, which are correlated each other. 

The equation of motion for protein atoms in water is {\em formally} akin to that of the Langevin equation for coupled harmonic oscillators in the continuum solvent, examined by Wang and Uhlenbeck long time ago. 
However, there exists a substantial and important difference.
Unlike the coupled set of harmonic oscillators, the "Hessian" included in the term corresponding to the Hookian-like restoring force in the new equation is identified as a variance-covariance matrix of the displacement vector, which is nothing but the second moment of the structural fluctuation. Since the fluctuation is taking place around the thermodynamic equilibrium, not just around a minimum of a (mechanical) harmonic potential, the "Hessian" matrix should be related to the second derivative of the free energy surface of protein, which of course includes the influence of solvent. 
    
    All those findings suggest that we are now at the position where we can explore the conformational fluctuation around a native state of protein, correlated with the relaxation of water density, since a method to evaluate the free energy surface and 
its first derivative of protein in water has been well established already based on the "3D-RISM/RISM" theory. 
It is not difficult to calculate the second derivative of the free energy surface from the first derivatives. 
    
 The finding further suggests even more practical applications related to the drug design. The 3D-RISM/RISM theory has been successfully applied to a variety processes of molecular recognition in protein, including drug binding. 
However, so far the application has been limited  to a fixed conformation of protein, which of course cannot take into account 
the effects of conformational fluctuation, such as the {\em induced fitting}.  
With the aid of the linear response theory, the new formulation  provides a foundation to evaluate the effect of conformational fluctuation in the process of molecular recognition.

\newpage
\appendix
\section{Calculation of  ${\bf B}_{\al,b} ({\bf k})$ }
\setcounter{equation}{0} 
We show that the correlation ${\bf B}_{\al,b} ({\bf k}) 
\equiv \big< \Del{\bf R}_{\al} \, \delta \rho^b_{{\bf k}}\big>$ 
vanishes in the homogeneous system. 
An explicit expression for ${\bf B}_{\al,b} ({\bf k})$ is  given by
\be
{\bf B}_{\al,b} ({\bf k}) \equiv \big< \Del{\bf R}_{\al} \, \del \rho^b_{{\bf k}}\big>
=\frac{1}{{\cal Z}_c} \int d \Gamma_c \,\, 
\Del{\bf R}_{\al} \, \sum_l  e^{i {\bf k} \cdot {\bf r}^b_l}  e^{-\beta {\cal U}}
\la{a1}
\ee
where $\Gamma_c$ denotes collection of the position variables only, and
$Z_c$ is the corresponding partition function defined as
$Z_c \equiv \int d \Gamma_c \exp \big( -\beta {\cal U} \big)$ with 
${\cal U}$ being the total potential energy of the system.  
Shifting  integration variables in (\ref{a1}) as 
${\bf r}^b_l \rightarrow {\bf r}^b_l +\Del{\bf R}_{\al}$ and
$\Del{\bf R}_{\beta} \rightarrow \Del{\bf R}_{\beta}+\Del{\bf R}_{\al}$ 
($\beta \neq \al$), 
one can separate the $\Del {\bf R}_{\al} $-integration as follows;
\be
\big<\Del {\bf R}_{\al} \del\rho^b_{{\bf k}}\big>=\frac{ \int d\Del{\bf R}_{\al}  \, 
\Del{\bf R}_{\al} \,  e^{i {\bf k} \cdot \Del {\bf R}_{\al}} }
{\int d \Del{\bf R}_{\al}} \cdot \frac{\int d{\bf R}^{N_u-1} \int d{\bf r}^{nN}
 \sum_l e^{i {\bf k} \cdot {\bf r}^b_l}  e^{-\beta {\cal U}'}}
{ \int d\Del{\bf R}^{N_u-1} \int d{\bf r}^{nN}   e^{-\beta{\cal U}'} }
\la{a2}
\ee
where the potential energy ${\cal U}' $ is obtained from ${\cal U}$ 
with the shift of the variables, and does not involve ${\bf R}_{\al}$.
We consider the first integration factor in (\ref{a2}). 
For simplicity we suppress the index $\al$.
Its $X$-component is given by
\bea
\frac{ \int d{\bf R}  \, X \,  e^{i {\bf k} \cdot {\bf R}} }{\int d{\bf R}}
&=& \frac{1}{L^3} \Big(\int_{-L/2}^{L/2} X e^{ik_1 X} dX \Big) 
\Big(\int_{-L/2}^{L/2}  e^{ik_2 Y} dY \Big)  \Big(\int_{-L/2}^{L/2}  e^{ik_3 Z} dZ \Big) \nonumber \\
&=& \frac{1}{L^3} \cdot \frac{2}{k_1} \big[\frac{1}{k_1} \sin \Big(\frac{k_1L}{2} \Big)
-\frac{L}{2}  \cos \Big(\frac{k_1L}{2}\Big)  \big]
\cdot \frac{2}{k_2} \sin \Big(\frac{k_2L}{2}  \Big) \cdot 
\frac{2}{k_3} \sin \Big(\frac{k_3L}{2}  \Big)
\la{a3}
\eea
Note that this integral vanishes in the thermodynamic limit $L\rightarrow \infty$ 
due to the oscillating term $e^{i{\bf k}\cdot {\bf R}}$.
The second integration factor in (\ref{a2}) remains finite. 
Therefore we obtain  $\big< X \rho_{{\bf k}}\big>=0 $ in the thermodynamic limit.
Since this will hold for the other components, we conclude that
\be
{\bf B}_{\al,b} ({\bf k}) = 0 
\la{a4}
\ee

\section{Calculation of $({\bf W}_{\al}, {\bf \Xi}^b_{\bf k}(t)) $}
\setcounter{equation}{0} 
Here we show that the memory matrix 
$({\bf W}_{\al}, {\bf \Xi}^b_{\bf k}(t)) $ vanishes.
We consider 
\be
\Big< {\bf W}_{\al}, {\bf \Xi}^b_{\bf k}(t) \Big>
=
\Big<{\bf W}_{\al} \,\, e^{it {\cal Q} {\cal L}} {\bf \Xi}^b_{\bf k}\Big>
=\Big<{\bf W}_{\al} \Big(1+t {\cal Q}i{\cal L}+\frac{t^2}{2!} 
{\cal Q}i {\cal L} {\cal Q} i {\cal L} +\cdots \Big) {\bf \Xi}^b_{\bf k}\Big>
\la{b1}
\ee
where ${\cal Q}=1-{\cal P}$.
For simplicity of notation,   we write ${\bf W}_{\al}$ and ${\bf \Xi}^b_{\bf k}$ 
from (\ref{eqn:3.31}) as 
\be
{\bf W}_{\al} ={\bf F}_{\al} + {\bf M}_{\al \beta} \Del{\bf R}_{\beta}, \qquad
{\bf \Xi}^b_{\bf k} = \dot {\bf J}^b_{\bf k}-{\bf C}_{be}({\bf k}) \,  \delta \rho^e_{\bf k}
\la{b2}
\ee
where the summation is implied for repeated indices, and the matrix 
${\bf M}\equiv k_BT {\bf L}^{-1}$ 
and ${\bf C}_{be}({\bf k}) \equiv i {\bf k} J_{bd}(k) \chi^{-1}_{de}(k)$.

We first consider the first term of (\ref{b1})
\bea
\Big<{\bf W}_{\al} {\bf \Xi}^b_{\bf k}\Big> &=&
 \Big<\big({\bf F}_{\al} 
+ {\bf M}_{\al \beta} \Del{\bf R}_{\beta}  \big) 
\big( \dot  {\bf J}^b_{\bf k}-{\bf C}_{be}({\bf k}) \,  \delta \rho^e_{\bf k}  \big) \Big>
\nonumber \\
&=&
\Big< {\bf F}_{\al} \dot  {\bf J}^b_{\bf k}\Big> - {\bf C}_{be}({\bf k})
 \Big< {\bf F}_{\al} \delta \rho^e_{\bf k} \Big>
+{\bf M}_{\al\beta} \Big< \Del{\bf R}_{\beta} \dot{\bf J}^b_{\bf k}\Big> 
 -{\bf C}_{be}({\bf k}) {\bf M}_{\al \beta} 
\Big<  \Del{\bf R}_{\beta} \del \rho^e_{\bf k} \Big>
\la{b3}
\eea
We already showed that the last two terms vanish ((\ref{eqn:3.22}) and Appendix A).
We now show that the two terms vanish as well.
Let us first consider the second term in (\ref{b3}). We have
\bea
\Big< {\bf F}_{\al} \delta \rho^e_{\bf k} \Big> 
&=& \frac{1}{Z}_c 
\int d\Gamma_c \,   \Big( \sum_l e^{i{\bf k} \cdot {\bf r}_l} \Big) {\bf F}_{\al}
e^{-\beta {\cal U}} 
= \frac{1}{Z}_c \int d\Gamma_c \,   \sum_l e^{i{\bf k} \cdot {\bf r}_l} 
\Big(-\frac{\partial {\cal U}}{\partial {\bf R}_{\al}}  \Big) 
e^{-\beta {\cal U}} \nonumber \\
&=& \frac{1}{Z}_c \int d\Gamma_c \,   \sum_l e^{i{\bf k} \cdot {\bf r}_l} 
 \beta^{-1} \frac{\partial} {\partial {\bf R}_{\al}}   e^{-\beta {\cal H}} 
 =0
 \la{b4}
\eea
where the last equality results upon doing the ${\bf R}_{\al}$-integration by parts.
Obviously we will have the same result for the first term, 
 $\Big< {\bf F}_{\al} \dot  {\bf J}^b_{\bf k}\big>=0 $,  since $\dot J^b_{\bf k}$ does not contain  ${\bf R}_{\al}$.
 Therefore we showed that 
\be
\Big< {\bf W}_{\al} {\bf \Xi}^b_{\bf k} \Big>=0.
\la{b5}
\ee

Next we consider  the second term in (\ref{b1}):
$\Big<{\bf W}_{\al} t {\cal Q} i{\cal L}{\bf \Xi}^b_{\bf k}\Big>
=\Big<{\bf W}_{\al} t \big(1-{\cal P} \big) \dot{\bf \Xi}^b_{\bf k} \Big>$.
We need to compute the term ${\cal P} \dot {\bf \Xi}^b_{\bf k} $.
From (\ref{eqn:2.3}), one can obtain
\bea
{\cal P} \dot {\bf \Xi}^b_{\bf k}&=&\frac{1}{k_BT} {\bf M}_{\al\beta} 
\Big< \Del{\bf R}_{\al} \, \dot {\bf \Xi}^b_{\bf k}\Big> \Del{\bf R}_{\beta} 
+
\frac{1}{M_{\al} k_BT} \Big< {\bf P}_{\al} \, \dot {\bf \Xi}^b_{\bf k} \Big> {\bf P}_{\al} 
\nonumber \\
&+&
 N^{-1}\chi^{-1}_{ad}({\bf k}) \Big<  \delta \rho^a_{-\bf k} \, \dot {\bf \Xi}^b_{\bf k} \Big> \delta \rho^d_{\bf k} 
+N^{-1} J_{ad}({\bf k}) \Big< {\bf J}^a_{-\bf k} \, \dot {\bf \Xi}^b_{\bf k} \Big> 
{\bf J}^d_{\bf k} \nonumber \\
&=& 
N^{-1}\chi^{-1}_{ad}({\bf k}) \Big<  \delta \rho^a_{-\bf k} \, \dot {\bf \Xi}^b_{\bf k} \Big> \delta \rho^d_{\bf k} 
+N^{-1} J_{ad}({\bf k}) \Big< {\bf J}^a_{-\bf k} \, \dot {\bf \Xi}^b_{\bf k} \Big> 
{\bf J}^d_{\bf k}
\la{b6}
\eea
where the last equality holds since 
$\Big< {\bf R}_{\al} \dot {\bf \Xi}^b_{\bf k} \Big>=
-\Big< {\bf P}_{\al} {\bf \Xi}^b_{\bf k} \Big>=0 $ 
and 
$ \Big< {\bf P}_{\al} \dot {\bf \Xi}^b_{\bf k} \Big>=0$.
Using (B.6), we have
\be
{\cal Q} \dot {\bf \Xi}^b_{\bf k}= (1-{\cal P}) \dot {\bf \Xi}^b_{\bf k}=
\dot {\bf \Xi}^b_{\bf k}-N^{-1}\chi^{-1}_{ad}({\bf k}) 
\Big<  \delta \rho^a_{-\bf k} \, \dot {\bf \Xi}^b_{\bf k} \Big> \delta \rho^d_{\bf k} 
-
N^{-1} J_{ad}({\bf k}) \Big< {\bf J}^a_{-\bf k} \, \dot {\bf \Xi}^b_{\bf k} \Big> 
{\bf J}^d_{\bf k}
\la{b7}
\ee
It is important to note that ${\cal Q} \dot {\cal R}_{\bf k} $ only involves the solvent coordinates and momenta. 
Then, using the orthogonality relations
 $\Big<{\bf W}_{\al} \delta \rho^d_{\bf k} \Big> =0$ and 
$\Big<{\bf W}_{\al} {\bf J}^d_{\bf k}\Big>=0$, 
one obtains
\bea
\Big<{\bf W}_{\al}  {\cal Q}i{\cal L} {\bf \Xi}^b_{\bf k} \Big>
&=& 
\Big<{\bf W}_{\al}  {\cal Q}\dot{\bf \Xi}^b_{\bf k} \Big>=
\Big< {\bf W}_{\al} \,\dot {\bf \Xi}^b_{\bf k} \Big>
=\Big< \Big( {\bf F}_{\al}+{\bf M}_{\al\beta}\Del{\bf R}_{\beta} \Big) \,
\dot {\bf \Xi}^b_{\bf k} \Big> \nonumber \\
&=&
\Big< {\bf F}_{\al} \dot {\bf \Xi}^b_{\bf k}\Big>
+{\bf M}_{\al\beta} \Big< \Del{\bf R}_{\beta} \,\dot {\bf \Xi}^b_{\bf k} \Big>=0
\la{b8}
\eea
where in the last line the first term vanish from the argument shown in (B.4), and 
we already showed the second term vanishes.
 
It is now clear that the repeated applications of ${\cal Q}i {\cal L}$ on 
${\bf \Xi}^b_{\bf k}$
 will never generate the solute-variable components, and hence
 $\Big< {\bf W}_{\al} ( {\cal Q}i {\cal L})^n {\bf \Xi}^b_{\bf k}\Big> =0$.
 Therefore we obtain the final result
 \be
 \Big< {\bf W}_{\al} e^{t {\cal Q}i{\cal L}}{\bf \Xi}^b_{\bf k}\Big>=0.
 \la{b9}
 \ee


\begin{thebibliography}{99}
\bibitem{jab}
The rigorous expression for $J_{ab}$ has been derived  by
the authors of Ref. \cite{yama-chong-hirata}.
\end{thebibliography}
\end{document}